\documentclass[a4]{article}

\usepackage{amssymb}

\usepackage{lineno}

\usepackage[figuresright]{rotating}

\usepackage{amsmath,amssymb,amsfonts}%
\usepackage{amsthm}%

\newcommand{\avg}[1]{\left< #1 \right>}

\usepackage{multirow}%
\usepackage{array} 

\usepackage[utf8]{inputenc}
\usepackage[T1]{fontenc}
\usepackage{lmodern}
\usepackage{textcomp}
\usepackage{amsmath}
\usepackage{amssymb}

\usepackage{xcolor}
\usepackage{soul}

\usepackage{graphicx}
\usepackage{epstopdf, epsfig}
\usepackage[lofdepth,lotdepth]{subcaption}

\usepackage{ifthen} 

\usepackage{enumitem}

\newcommand{\nk}[1]{{\small\setlength{\baselineskip}{1.25em}%
    {\it [{\color{blue} nk: #1}]}}}
\newcommand{\cc}[1]{{\small
    {\it [{\color{green} cc: #1}]}}}
\newcommand{\mm}[1]{{\small
    {\it [{\color{orange} mm: #1}]}}}

\NewCommandCopy{\oldcite}{\cite}
\renewcommand{\cite}[2][]{%
  {%
    \oldcite{#2}%
  }%
}

\renewcommand{\nk}[1]{}
\renewcommand{\cc}[1]{}
\renewcommand{\mm}[1]{}

\newcommand{\pder}[2]{\frac{\partial#1}{\partial#2}}

\newcommand{\air}{}
\newcommand{\wat}{^{(w)}}
\newcommand{\Uwind}{{\mathcal{U}}}

\newcommand{\coeff}[1]{\ensuremath{\mathcal{#1}}}

\newcommand{\Hcoeff}{\coeff{H}}
\newcommand{\Xcoeff}{\coeff{X}}

\newcommand{\zo}{z_{\emptyset}}  

\newcommand{\ie}{\textit{i.e.} }
\newcommand{\eg}{\textit{e.g.} }

\newcommand{\sss}{\scriptscriptstyle}

\usepackage[subdued]{mathastext}
\MTfamily{augie}
\Mathastext[augie]

\usepackage{mdframed}

\usepackage{marginnote}

\usepackage{etoolbox}
\usepackage{environ}

\mdfdefinestyle{tostyle}{%
  topline=false,
  bottomline=false,
  skipabove=0cm,
  skipbelow=0cm,
  leftmargin=-10pt,
  rightmargin=-10pt,
  innertopmargin=0pt,
  innerbottommargin=0pt,
  linewidth=0.2em}

\newtoggle{todo}
\definecolor{todocolor}{rgb}{0.3, 0.8, 0.2}

\newcommand{\todomargin}{!!}
\newcommand{\thetodomargin}{\bf\color{todocolor}\marginnote[{$\Rightarrow$}\todomargin]{$\Leftarrow$ \todomargin}}%

\NewEnviron{todoenv}{\thetodomargin\BODY}

\surroundwithmdframed[style=tostyle,linecolor=todocolor,fontcolor=todocolor]{todoenv}

\usepackage{changepage}

\newtoggle{tothink}
\definecolor{tothinkcolor}{rgb}{0.8, 0.2, 0.2}

\newcommand{\tothinkmargin}{??}
\newcommand{\thetothinkmargin}{\color{tothinkcolor}\marginnote[{$\Rightarrow$}\tothinkmargin]{$\Leftarrow$ \tothinkmargin}}%

\NewEnviron{tothinkenv}{%
  \small%
  \setlength{topsep}{0cm}%
  \setlength{parskip}{0cm}%
  \setlength{partopsep}{0cm}%
  \thetothinkmargin\BODY}
\surroundwithmdframed[style=tostyle,linecolor=tothinkcolor,fontcolor=tothinkcolor]{tothinkenv}

\newtoggle{torecall}
\definecolor{torecallcolor}{rgb}{0.2, 0.2, 0.8}

\newcommand{\torecallmargin}{rcll
  !}
\newcommand{\thetorecallmargin}{\color{torecallcolor}\marginnote[{$\Rightarrow$}\torecallmargin]{$\Leftarrow$ \torecallmargin}}%

\NewEnviron{torecallenv}{
  \setlength{\itemsep}{0cm}%
  \setlength{\topsep}{0em}%
  \setlength{\parskip}{0.5em}%
  \setlength{\partopsep}{0cm}%
  \thetorecallmargin\BODY}

\surroundwithmdframed[style=tostyle,linecolor=torecallcolor,fontcolor=torecallcolor]{torecallenv}

\newtoggle{toread}
\definecolor{toreadcolor}{rgb}{0.7, 0.1, 0.7}

\newcommand{\toreadmargin}{rd!}
\newcommand{\thetoreadmargin}{\color{toreadcolor}\marginnote[{$\Rightarrow$}\toreadmargin]{$\Leftarrow$ \toreadmargin}}%
\newcommand{\toreadbody}[1]{%
  \iftoggle{toread}{{%
    \thetoreadmargin
    {\noindent\color{toreadcolor}%
      \textit{[{$\to$ }#1]}%
    }}%
  }{%
  }
}

\newcommand{\toread}[1]{\iftoggle{toread}{\toreadbody{#1}}{}}
\NewEnviron{toreadenv}{%
  \setlength{\itemsep}{0cm}%
  \setlength{\topsep}{0em}%
  \setlength{\parskip}{0.5em}%
  \setlength{\partopsep}{0cm}%
  \thetoreadmargin\BODY}

\surroundwithmdframed[style=tostyle,linecolor=toreadcolor,fontcolor=toreadcolor]{toreadenv}

\toggletrue{todo}
\toggletrue{tothink}
\toggletrue{torecall}
\toggletrue{toread}

\togglefalse{toread}

\usepackage{titlesec}

\usepackage{url}

\begin{document}



\title{A plea for Miles' approach to wind-wave interaction} 

\author{
  \parbox{\textwidth}{
    Norbert Kern$^1$\footnote{norbert.kern@umontpellier.fr} ,
    Miguel M. Manna$^1$\footnote{miguel.manna@umontpellier.fr} ,
    Frédéric Bouchette$^2$\footnote{frederic.bouchette@umontpellier.fr} ,
    Christophe Chaubet$^1$\footnote{christophe.chaubet@umontpellier.fr}
    \\[0.5em]
    1 - L2C, Université Montpellier, CNRS, Montpellier, France\\
    2 - Université de Montpellier, INRIA, Montpellier, France
    }
}

\maketitle

\begin{abstract}
  Although the phenomenon of wave growth under the effect of wind still holds many open questions, the approach initiated by Miles \cite{Miles:1957} is a cornerstone of the current comprehension. This theory is based on physical arguments, keeping phenomenological input to a minimum, but it can be perceived as intimidating, especially in view of concrete applications. Here we argue that Miles' results can be summarised in terms of a simple functional form, which provides an easy-to-use expression to work with. We furthermore show that Miles' predictions do not, as is often stated, underestimate growth rates, once the presence of a whole wave spectrum is accounted for. Finally, we illustrate that the generalisation of Miles' original result to complex hydrodynamic situations can be quite straightforward. Our hands-on summary  should make it heretofore very easy to exploit Miles' approach in practice.
\end{abstract}







\section{Introduction}

\subsection{Context}
Miles' approach is a cornerstone for understanding the interaction between wind and the sea surface, and thus the growth of surface gravity waves.
Put forward by Miles in 1957 \cite{Miles:1957}, this seminal work addresses the question of the mechanism itself which makes ocean waves grow under the effect of wind. The approach consists in a full hydrodynamic analysis, stating the problem in terms of a two-fluid model (water and air, separated by a free interface on which a sinusoidal perturbation propagates).
Miles' analysis addresses the growth, but not the initial formation of surface waves: for this, the approach by Phillips \cite{Phillips:1958} has been seen to be more relevant, as it reproduces the initially linear growth of amplitude with time, which has recently been validated in laboratory experiments \cite{Paquier:2015,Paquier:2016,Paquier:thesis:2017}.
For the ensuing  phase of exponential growth, Miles has established expressions for several growth coefficients.
Initially, this was formulated for the simplified situation of an idealised ocean (very deep, incompressible and inviscid water, without any currents).
Nevertheless, the {\it method} put forward by Miles is not limited to this condition, and has been extended later by Miles himself \cite{Miles:1959,Miles:1959a,Miles:1960,Miles:1962,Miles:1967}. Our contribution, which homogenises these studies, is precisely about making Miles' method simple to use and extend beyond the idealised reference case.

Refinements to Miles'approach have been produced over the years \cite{Miles:1959,Miles:1959a,Miles:1960,Miles:1962,Miles:1967,Valenzuela:1976} and its limitations have been pointed out: see \cite{Krasitskii:1978,Riley:1982,Belcher:1993,Janssen:book}, to cite but a few authors. A numerical approach to calculating the growth rates as a function of wind strength has been put forward \cite{Conte:1959} and revisited more recently \cite{Beji:2004}. Extensions to less restrictive hydrodynamic assumptions have been provided to account for finite water depth \cite{Montalvo:2013a,Montalvo:2013b,Montalvo:thesis:2014}, for a shear flow in the water domain \cite{Kharif:2020,Kern:2021}, and for the bulk viscosity of water \cite{Miles:1959a,Benjamin:1959,Chaubet:2024}.
 \toread{Miles:1993 also refers to resonance with Tollmien-Schlichting waves, see Miles:1962: this is again about the Reynold's stress tensor}
 A framework for generalising the approach to yet more general hydrodynamic conditions has been put forward recently \cite{arxiv}.
 Carefully designed experiments have been undertaken to study in particular the impact of finite water depth \cite{Branger:2022} and of viscosity \cite{Paquier:2015,Paquier:2016}.

Miles' approach is particularly interesting in that it prioritises physical arguments over phenomenological ad-hoc assumptions. In principle  the refinements since  Miles' seminal article \cite{Miles:1957} make this approach a solid framework, suitable also for use in near-shore conditions - which is of course a major stake in modelling ocean waves. However, we are not aware of recent developments applying this by now rather complete theory in practical studies or predictive simulations.

However, despite the clear-cut spirit of Miles' approach, it appears to be under-exploited, for which there are at least three reasons:
  (i) the framework can appear intimidating, and may give the impression of being difficult to use, in particular in view of the need for numerical work;
  (ii) Miles' results for the growth rates are typically considered to underestimate the growth rates with respect to what has been observed, on the ocean and in laboratory experiments \cite{Plant:1982};
  (iii) adapting it to other hydrodynamic situations is expected, at first sight, to be  complex and lengthy.

\toread{Miles:1993 avait proposé une autre explication en termes de 'eddy viscosity'}
  Here we show that these points no longer constitute serious obstacles: Miles' approach is much simpler to use than one might think, and it provides accurate results.
After an overview, a complete section will be dedicated to each if these points.

\section{Miles' approach:  an overview}

Miles has developed his arguments in the simplest possible situation, an infinitely deep ocean of incompressible, inviscid water without any currents, ignoring interfacial effects such as capillarity. We first focus on this specific situation, which will serve as a reference case throughout, to set up the mathematical model.

\subsection{Parametrising the wave dynamics}

\begin{figure}
  \center
  \label{fig:sketch:setup}
    \includegraphics[width=0.5\textwidth]{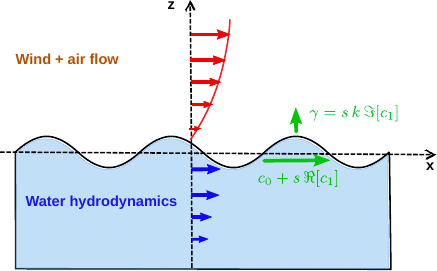}
  \caption{
    \label{fig:sketch}
    Schematic representation of Miles' two-domain model.
    The air domain consists of the (imposed) wind profile as well as the perturbations of the air flow caused by the wave. Similarly, the water domain accounts for hydrodynamics, including the perturbation due to the wave. Via the coupling at the interface one derives the modification $s  c_1$ of the celerity attributable to the air, see Eq.~(\ref{eq:c0+sc1})) which directly yields the wind-generated amplitude growth rate $\gamma$, see Eq.~(\ref{eq:wave}).   
  }
\end{figure}

A wave propagating in the positive $x$ direction (see Fig.~\ref{fig:sketch}) is parametrised as
\begin{equation}
  \label{eq:wave}
  \eta \, e^{ik(x-ct)} = \eta \, e^{ik(x-\Re[c]t)} \, e^{\gamma t}
  \ ,
  \qquad\mbox{with} \qquad
  \gamma := k \, \Im[c]
\end{equation}
where $\Re[.]$ and $\Im[.]$ refer to the real and imaginary parts of the argument. $\eta$ is the amplitude of the surface elevation of the air-water interface, $k$ is the (norm of the) wave vector, $\Re[c]$ is the propagation velocity, and the relation specifying $c(k)$ as a function of $k$ is referred to as the {\it dispersion relation}.

Determining the growth rate for the wave amplitude thus amounts to finding the imaginary part of the celerity $c(k)$ in the situation where air flow is present, with $U_{ext}$ imposing the external flow for the desired wind profile.

\subsection{Wind profile and roughness length}
Before discussing the approach by Miles in detail, we point out that we shall use a logarithmic wind profile for illustration:
\begin{equation}
  \label{eq:windprofile:log}
  U_{ext}(z) = {\mathcal U} \, \ln \frac{z}{\zo} \ = \frac{u^*}{\kappa} \, \ln \frac{z}{\zo} \ ,
\end{equation}
where  $\zo$ is the {\it roughness length} and we refer to  ${\mathcal U}$ as the {\it wind strength}, which is related to the more commonly used {\it friction velocity} $u^*$ as $\Uwind = \frac{u^*}{\kappa}$, with $\kappa=0.41$ being the Von-Kármán constant.
The roughness length $\zo$  is usually determined from Charnock's relation
\begin{equation}
  \label{eq:charnock}
  \zo
  = \alpha_{\sss CH} \, \frac{\left(u^*\right)^2}{g}
  = \Omega_{\sss CH} \, \frac{\Uwind^2}{g}
  \ ,
\end{equation}
where $\Omega_{\sss CH} = 3 \cdot 10^{-3}$ is a rather consensual value, already used by Miles. It thus refers back to the original {\it Charnock constant} $\alpha_{\sss CH}= 0.0178$ as $\Omega_{\sss CH}= \kappa^2 \, \alpha_{\sss CH}$ \cite{Charnock:1955}.
Moderate modifications to these constants have been explored with ultimately little impact \cite{Beji:2004}.

This logarithmic profile is widely used, and is validated in experiments \cite{Branger:2022}. Nevertheless, different profiles have been considered, for example by Beji \& Nadaoka \cite{Beji:2004}, and can be treated by the same method.
Throughout this manuscript we use the profile given by Eqs.~(\ref{eq:windprofile:log}) and (\ref{eq:charnock}), which is based on a description of eddies on all length scales in terms of the mixing length model \cite{Monin:1959,Landau:book:FluidMechanics,Prandtl:1925,Prandtl:1949}. However, importantly, the approach could easily be adapted to a different profile if required.

\subsection{Miles' aproach}

In a nutshell, Miles has suggested the following strategy:
(i) the air being much less dense than water, $s:=\rho\air/\rho\wat \ll 1$ is a small parameter; thus the effect of air flow (and wind in particular) on the wave can be evaluated for the initial wave (no feedback loop between air and water domain is required); (ii) the air pressure perturbation provoked by the wave can be quantified and is seen to obey a differential equation known as the {\it Rayleigh equation}; 
(iii) from a solution to the Rayleigh equation one can thus deduce the modification of the celerity, due to the presence of air and wind, and in particular its complex part which yields the growth coefficient via Eq.~(\ref{eq:wave}).

Thinking of the density ratio $s=\rho\air/\rho\wat \ll 1$ as an expansion parameter, all quantities are obtained starting from the 'airless' case, $s=0$ (for which the air is replaced by a constant pressure field) by adding a first-order correction in $s$. For example, for the celerity 
\begin{equation}
  \label{eq:c0+sc1}
  c = c_0 + s \, c_1 \ .
\end{equation}
For Miles' reference conditions $c_0=\sqrt{g/k}$. Determining the coefficient $c_1$ is therefore the practical goal since, by the virtue of Eq.~(\ref{eq:wave}), $\Im[c_1]$ directly yields the amplitude growth coefficient due to wind.

\subsection{Intrinsic limitations of Miles' approach vs.  choice of simplified hydrodynamic conditions}
A major appeal of the strategy put forward by Miles is clearly that it is constructed on physical arguments, rather than ad-hoc parameters or phenomenological hypotheses. This also means that the limitations of the approach are straightforward to state:
\begin{itemize}
\item
  The wave amplitude is taken to be 'small' on the scale of the wave ($k \eta \ll 1$). For ocean waves this is true for long stretches of time, since the growth is slow. But the theory must undoubtedly fail at last, due to the unbounded growth it predicts (in particular, breaking waves constitute the ultimate failure of this assumption, and are entirely outside of the scope of this description).
\item
  Both water and air are taken to be incompressible, and the air is inviscid, which constitutes a very reasonable starting point for ocean waves.
\item
  The water is not necessarily inviscid but its viscosity must remain 'weak' (see \cite{arxiv} for a detailed consideration. This condition leads to requiring $\nu \, k / \Re[c_0] \ll 1$, which is true for typical ocean waves but also in many laboratory experiments, such as those by \cite{Paquier:2015,Paquier:2016}.
  
\item
  The density ratio $s:=\rho\air/\rho\wat$ between air and water density must be small (as is the case, $s \simeq 10^{-3}$, for water waves on earth, but this condition also holds for more exotic oceans on exo-planets \cite{Schneck:2026}).
  
\item
  Finally, the effects of a boundary layer in the air domain are taken to be captured through the aforementioned {\it roughness layer}.
\end{itemize}

These restrictions, inherent to the approach proposed by Miles, are crucial for mathematical feasibility.
To the contrary, it is worth insisting again that Miles' approach is {\it not} restricted to a specific wind profile.

\subsection{Miles' seminal results}

Miles' initial publication \cite{Miles:1957} has focused on the simplest choice of hydrodynamics: entirely inviscid water and air, infinite water depth, no water currents, and absence of interfacial effects such as surface tension. 
In contrast to the intrinsic assumptions listed above, these conditions can (and have since been) relaxed, and we will return to this point in section \ref{sec:hydro}. We shall label results specific to these reference conditions with a subscript 'M'.

Miles' seminal paper  has established an expression for the growth rate $\gamma_{\sss M}$, in these hydrodynamic conditions.
Once stated formally, the expression for $\gamma_{\sss M}$ is then hard to evaluate, since it implicitly requires solving the Rayleigh equation. Miles has then discussed the salient predictions, using rather delicate analytical approximations, and later on a numerical approach \cite{Conte:1959}.
More recently, Beji \& Nadaoka \cite{Beji:2004} have suggested a  detailed numerical procedure for evaluating $\gamma_{\sss M}$ for given wave vector $k$ and wind strength $\mathcal{U}$.

\subsection{The wave age}

The resulting growth coefficient is often represented as a function of the ratio $c/\Uwind$, or its inverse \cite{Miles:1957,Beji:2004,Montalvo:2013a,Montalvo:2013b,Montalvo:thesis:2014,Kern:2021,Chaubet:2024}. This is related to a parameter  $\theta$ sometimes used in field work, known as the {\it  wave age}.
Specifically, in experimental work $\theta$  is defined as the ratio of the propagation velocity $\Re[c]$ to wind speed $\Uwind$ at the peak of the wave spectrum, $\theta := \Re[c_p] / {\mathcal U}$, see for example \cite{Janssen:book,Plant:1982,Donelan:1992,Young:1996}.
Since here we are dealing with a single wave length this simplifies to
\begin{equation}
  \theta := \frac{\Re[c]}{{\mathcal U}} 
  \ ,
\end{equation}
which we will refer to as the {\it monochromatic wave age}.

Note that one must in principle distinguish the celerity of the actual wave  ($c$) from that of the reference wave in the 'airless' system ($c_0$, at $s=0$). Although the distinction is conceptually important (as is discussed rigorously in \cite{arxiv}), neglecting the difference only produces much smaller corrective terms (higher powers in $s \approx 10^{-3}$), which are already neglected in the approach. To simplify the discussion we can therefore refer indiscriminately to both
  $
  \theta = \frac{\Re[c]}{\Uwind}
  $
  and
  $
  \theta_0= \frac{\Re[c_0]}{\Uwind}
  $ 
as '{\it the wave age}'.
In Miles' reference case, $\Re[c_{0,M}] = c_{0,M} = \sqrt{g/k}$.

Some of the cited works use alternative definitions, such as an equivalent, $\theta^*$ based on the friction velocity $u^*$. Another variant is based on  $U_{10}$, the wind speed at a height of 10\,m  as a characteristic wind speed, defining the wave age as the ratio $c(k_p)/U_{10}$.
When required, conversions are easily made. For example, for a logarithmic wind profile, the correspondence is
$U_{10} = U \, \kappa/\sqrt{C_{10}}$ in terms of the drag coefficient $C_{10}$, given by the  empirical relation  $C_{10} = (0.75 + 0.067 \, U_{10}) \times 10^{ -3}$, as given by Young\cite{Young:book}.

\subsection{The wave age as key  to Miles' approach}

Starting with Miles \cite{Miles:1957}, many authors have chosen to present results as a function of the wave age.  This is made possible because, for a given velocity profile, the expression for $\gamma_{\sss M}$ is seen to be a function of the ratio $c/{\mathcal U}$ alone.
The same holds for a different, rescaled growth coefficient
\begin{equation}
  \label{eq:beta:def}
  \beta := \frac{2}{s} \, (\gamma_{\sss M}/\omega) \, \theta^2
  \qquad\mbox{(simplified hydrodynamic conditions)}
  \ ,
\end{equation}
directly related to the pressure perturbation due to the wave, which is often preferred in experimental or observational work.
Despite the fact that converting between the $\gamma$ and $\beta$ growth parameters is of course straightforward, considering $\beta$  is particularly convenient for our purposes, since it turns out \cite{arxiv} that the pressure coefficient $\beta$ is not restricted to the simplified hydrodynamic conditions: it is seen to be truly independent of the hydrodynamics, and therefore characterises wind-wave interaction as such, whereas this is not true for $\gamma$, which also comprises the hydrodynamic response of the water \cite{arxiv}.

The behaviour of $\beta$, as characterised by Miles and deduced by numerical work following Beji \cite{Beji:2004}, is represented in Fig.~\ref{fig:fit:full}. In essence, the wave growth coefficient is approximately constant for a moderate wave age up to five, but vanishes as the wind speed drops below roughly one tenth of the wave speed ($\theta_0 > 10$), which corresponds roughly\footnote{
  Using $C_{10} = 10^{-3}$ in the expression for $U_{10}$, the value $\theta_0=12$, which is where $\beta$ vanishes, is seen to correspond to $c/U_{10} \approx 0.93$
}
to $c > U_{10}$: once the wave outruns the wind (measured at a height of 10 m), the energy transfer from wind to the wave essentially fails.
Note also that this theory does not claim to be valid for small wave ages, say $\theta < 1$: very strong winds would require a more refined analysis \cite{Miles:1993}.

\subsection{Outline} 
We point out  that the discussion has so far been limited to the simplified hydrodynamic conditions adopted by Miles in his initial analysis, which greatly limits the applicability of the results. We return to this point below (section \ref{sec:hydro}), where we argue that generalising to more complex hydrodynamic situations is much easier than one might apprehend. However, we first address the question of casting Miles' result into a practically convenient form, which is the goal of the next section. 

\section{A convenient expression for Miles' growth coefficients}

Miles' approach as outlined above retains some technical challenges, such as numerically implementing the method proposed by Beji \& Nadaoka \cite{Beji:2004}, with the assorted convergence issues. No explicit analytical solution is available, and  working with tables is awkward. Thus one is left with fitting a function to obtain a workable solution for all practical purposes.

Finding ad-hoc functions for a fit is not an easy task, if the expected functional dependence is unknown. Here we start by observing the threshold-like behaviour in $\beta(\theta)$ already pointed out above, which suggests that a hyperbolic tangent will probably be useful. We have then improved on the resulting fit, rough but promising, by modulating the argument of the $\tanh$ function and fitting as
\begin{equation}
  \label{eq:fit:full}
  \beta(\theta) = 1.74 \, \left( 1-\tanh\left[ \frac{\theta-8.01}{2.28}  \,  \left(1+e^{\theta-11.83} \right) \right] \right)
  \ .
\end{equation}

\begin{figure}
  \includegraphics[width=\textwidth]{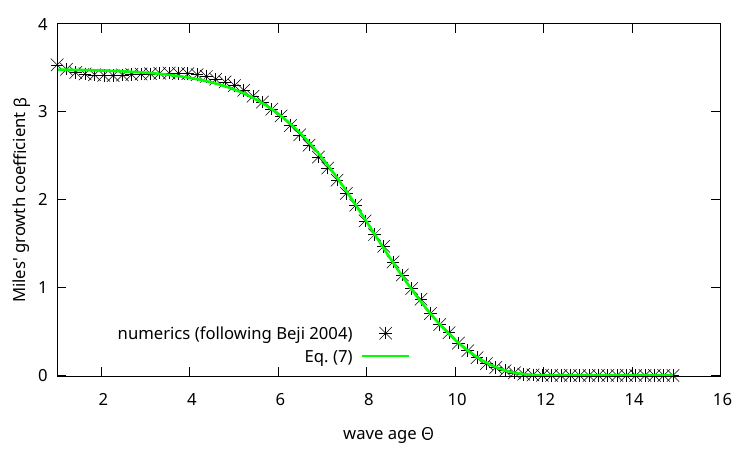}
  \put(-225,140){\includegraphics[width=0.45\textwidth]{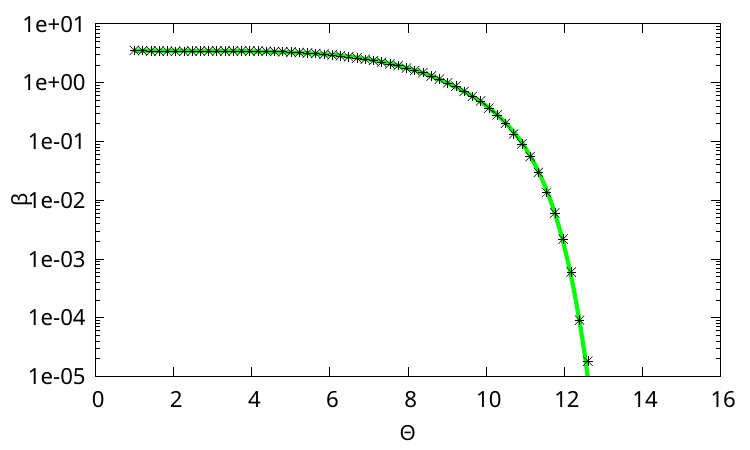}}
  \caption{\label{fig:fit:full}
    Plot of Miles' growth coefficient $\beta$, determined  according to the numerical approach due to Beji \& Nadaoka (black crosses), juxtaposing the expression proposed in Eq.~(\ref{eq:fit:full}) (green line). The agreement is very good both on a linear scale (main plot) and on a log-lin representation (inset, showing the same data). Note that a simple step function is not sufficient to achieve this.
  }  
\end{figure}

A plot of Eq.~(\ref{eq:fit:full}) is shown in Fig.~\ref{fig:fit:full}, juxtaposing it to numerical data obtained using Beji \& Nadaoka's method.
Recalling that the domain of small $\theta$ is out of scope of this theory (see above) one may conclude that the above function is largely sufficient for any practical purposes. Importantly, the fit matches both on a linear and on a logarithmic scale (see Fig.~(\ref{fig:fit:full}), which sets it apart from other ad-hoc formulas.

Having a reliable fitting function is not only convenient but, first and foremost, avoids performing a numerical analysis for every single wavelength for which the growth rate is required. This is significant since it circumvents any subtleties in implementing a numerical procedure, and is computationally efficient: evaluating Eq.~(\ref{eq:fit:full}) is essentially instantaneous, and does not limit performance with respect to other numerical calculations involved. These are key requirements for making extensive use of such refined predictions in large-scale oceanographic computations.

The essential result of this result is the fit (Eq.~(\ref{eq:fit:full})), which yields the amplitude growth coefficient by re-writing Eq.~(\ref{eq:beta:def})  as
\begin{equation}
\frac{\gamma_{\sss M}}{\omega} = \frac{s}{2} \, \frac{\beta(\theta)}{\theta^2}   
  \ .
\end{equation}
The major importance of this result will emerge with the final section, where we show that the expression for $\gamma_{\sss M}$ also gives access to the growth coefficient beyond Miles' simplified reference conditions including  much more general hydrodynamic situations in the water domain, still not requiring any numerical work at all.
For now, we address the question of confronting Miles' predictions to data.

\section{Miles' predictions do not underestimate growth}

\begin{figure}  
  \includegraphics[width=\textwidth]{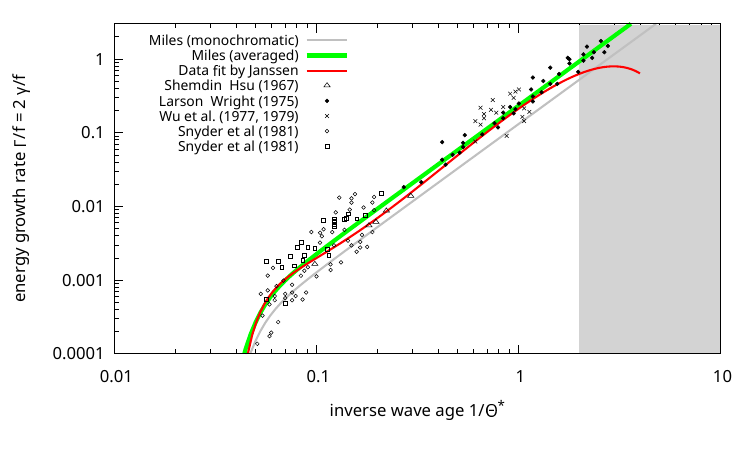}
  \caption{\label{fig:realdata}
    Energy growth coefficient $\Gamma=2\gamma$, non-dimensionalised as $\Gamma/f$ with $f=\omega/(2\pi)$ and presented as a function of the inverse wave age $1/\theta^{*} = 0.41/\theta$ (following the historical way of plotting, see \cite{Plant:1982}, but note that this reference uses the letter $\beta$ for a growth rate $\gamma$). The linear range $\gamma \sim (1/\theta^*)^2$ corresponds to the plateau zone in $\beta(\theta^*)$.
    Data points are from \cite{Plant:1982}, compiled from various observational studies.
    The fine red  line is the ad-hoc function suggested by Janssen \cite{Janssen:1991,Janssen:book} in order to fit the data.
    The fine gray line represents the original prediction by Miles, based on a monochromatic wave, taken here to correspond to the peak in the spectrum. It clearly underestimates the data, roughly by a factor of 2.
    The bold green line is the prediction for a spectrum, where the growth coefficient of each contributing mode is still determined from Miles' $\gamma(\omega)$. However, a spectrally weighted average (Eq.~(\ref{eq:spectrum:average}))
    using the wave spectrum due to Pierson \& Moskovitz, Eq.~(\ref{eq:spectrum:PM}) clearly achieves a very good match with the experimental data.
    The shaded region is a reminder that Miles' approach cannot be expected to work for very small wave ages.
    A detailed description of how the graphs in this figure are constructed is provided in \ref{app:log4}.
   }
 \end{figure} 

Having argued that Miles' results can be applied with ease,  the predicted wave growth coefficients still have to be confronted to measured data in order to validate, or to refute, this approach. This has already been addressed by Plant {\it et al} \cite{Plant:1982} as early as 1982, who have established a comparison based on data from several observational studies. It suggests that Miles' prediction is qualitatively correct but systematically underestimates the growth coefficients, usually by a factor of 1.5 or 2 \cite{Komen:1984,Janssen:book}. This is presented as a shortcoming of Miles' approach.

Even though deviations are entirely expected, due to the uncertainty inherent in the measurements used for  validation \cite{Young:1996,Villefer:2021,Branger:2022}), one would still hope to explain why the growth coefficients appear to be consistently underestimated by Miles' approach. 
Here we provide a new element for this explanation. Usually, measurements are not performed for a single wave length, but in a flume or in a field of waves.
Nevertheless, when a growth coefficient is measured, it is typically attributed to the principal mode, {\it i.e.} the peak mode $\omega_p$ in the wave spectrum, where $k_p$ and $\omega_p$ are of course linked by the dispersion relation.

Miles' approach should therefore not be used to simply determine the growth coefficient $\gamma_{\sss M}(\omega_p)$ attributed to  this single mode $k_p$.
Rather, we should emulate the spectral averaging which  occurs implicitly in observations.
To do so, we note that the spectrally weighted average growth coefficient, the one which corresponds to a spectrum of waves parametrised as $S(\omega)$, can be expressed as
\begin{equation}
  \label{eq:spectrum:average}
  \avg{\frac{\gamma}{\omega}} = \frac{\int_{0}^{\infty} \frac{\gamma(\omega)}{\omega} \, S(\omega) \, d\omega}{\int_{0}^{\infty} S(\omega)  \, d\omega}
  \ .
\end{equation}
The resulting spectrally weighted average value $\avg{\gamma/\omega}$ is then attributed to the peak frequency $\omega_p$.

For the sake of illustration we work with the JONSWAP spectrum \cite{jonswap:1973} with an enhancement factor of 1:
\begin{equation}
  \label{eq:spectrum:PM}
S(\omega) = \frac{\alpha g^2}{\omega^5} \, e^{- \frac{5}{4} \, \left(\frac{\omega_p}{\omega}\right)^4}
\ ,
\end{equation}
where  $\alpha=8.10 \cdot 10^{-3} $ and  $\omega_p$ is a parameter controlling the position of the peak of the spectrum. This is equivalent to the wave spectrum established by Pierson \& Moskovitz \cite{Pierson:1964}, but with an adjustable $\omega_p$. The spectral generalisation of the energy growth coefficient $\Gamma$, related to $\gamma$ as
\begin{equation}
  \frac{\Gamma}{\omega} = 2 \, \frac{\gamma}{\omega}
  \ ,
\end{equation}
can then be determined by performing the integration in Eq~(\ref{eq:spectrum:average}) numerically. This is represented in Fig.~(\ref{fig:realdata}), which also shows data points (symbols, corresponding to various experimental data sets, taken from \cite{Plant:1982}, as well as the ad-hoc fit to this data formerly proposed by Janssen \cite{Janssen:1991,Janssen:book}. Where Miles' prediction based on a monochromatic wave (grey line) is clearly low with respect to the data, introducing an averaged spectral wave growth  coefficient $\avg{\Gamma}$, given by Eq.~(\ref{eq:Gamma}), shifts it right onto the data. In  \ref{app:averaged} we show that, for the Pierson-Moskovitz spectrum, this shift on the logarithmic scale corresponds to a correction by a factor of $\sqrt{5\pi/4}$ to the growth coefficient, which can be attributed directly to the averaging, so that
\begin{equation}
\avg{\frac{\Gamma}{\omega}} = \sqrt{\frac{5 \pi}{4}} \,  s  \, \frac{\beta(\theta_p)}{\theta_p^2} \ .
\end{equation}
Note that $\sqrt{5 \pi/4} \approx 1.98$, which thus indeed explains a deviation by a factor of roughly 2.

\section{Extending Miles' results to more complex hydrodynamic situations}
\label{sec:hydro}

\begin{figure}
  \centering
  \includegraphics[width=0.7\textwidth]{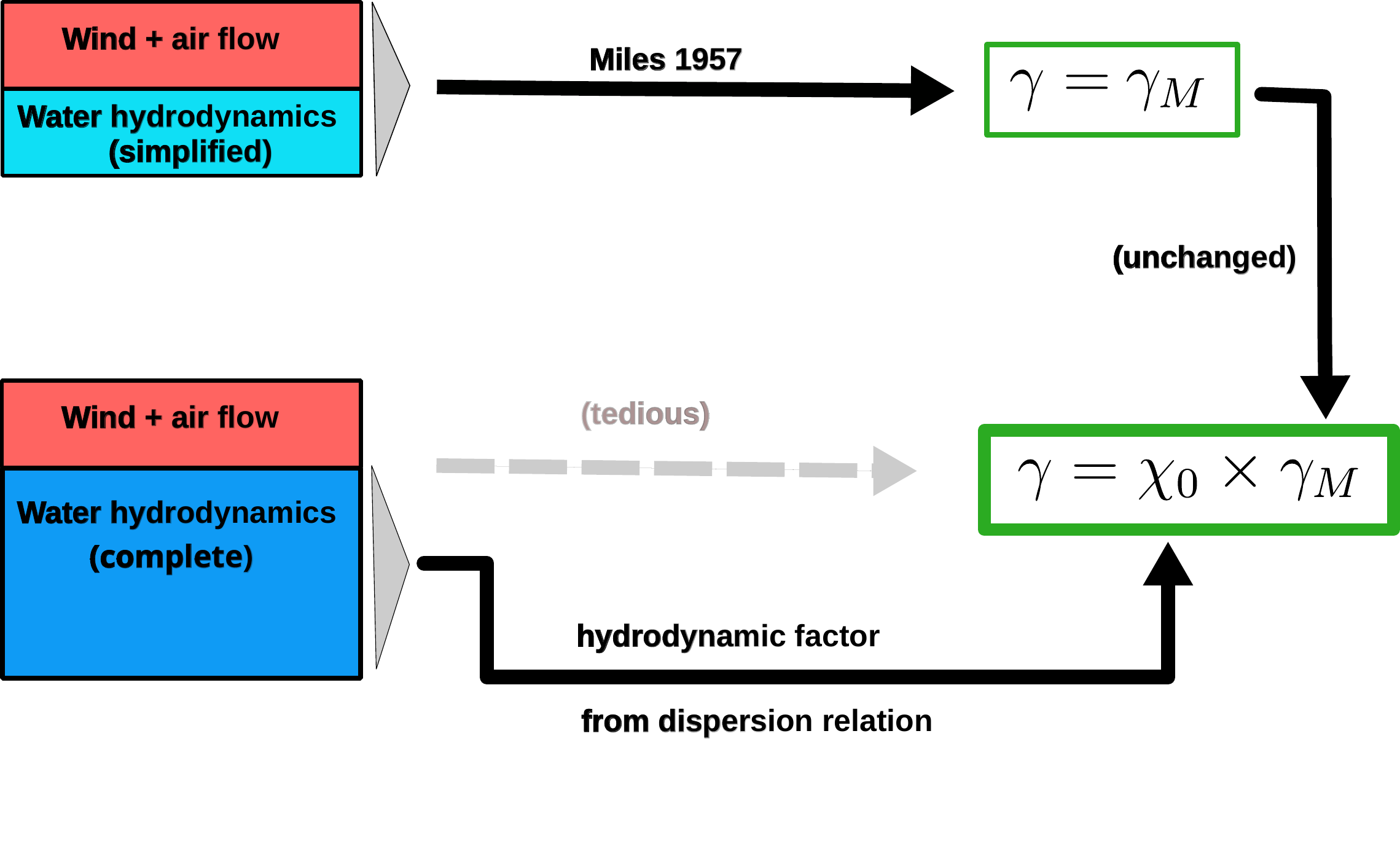}
  \caption{
    \label{fig:diagram:procedure}
    Diagram summarising the approach. Miles' seminal article has established the growth coefficient $\gamma_{\sss M}$ for the specific case of simplified hydrodynamics (infinitely deep still ocean without viscosity or capillarity). Our approach generalises $\gamma$ to general hydrodynamic conditions. The extra hydrodynamic effects are captured by a simple factor $\chi_0$, which can be calculated in a very straightforward way, based only on the dispersion relation (Eq.~(\ref{eq:c0(k)})) of a wave in the absence of air (and wind).
  }
\end{figure} 

Although Miles' seminal article was intentionally focused on the simplified hydrodynamic situation indicated above, it has since been extended beyond these reference conditions, accounting for other factors such as finite water depth \cite{Miles:1993,Montalvo:2013a,Montalvo:2013b,Montalvo:thesis:2014}, vorticity \cite{Kharif:2020,Kern:2021} or viscosity \cite{Miles:1993,Chaubet:2024}.

Rather intriguingly though, this can be done in a very generic way, which ultimately recasts the problem of coupling to the wind in modified hydrodynamic conditions into simple algebraic considerations.  Full details of the framework \cite{arxiv} are of no importance here, but we give a brief overview of these developments, as well as detailed examples, in \ref{app:framework}. They illustrate that these extensions constitute a powerful yet painless way to adapt Miles' approach to other, more realistic hydrodynamic situations. The approach is schematically captured in Fig.~\ref{fig:diagram:procedure}.

Here we simply paraphrase the approach to show that it makes much more general hydrodynamic situations tractable based on Eq.~(\ref{eq:fit:full}). We start from the formula for the dispersion relation as a function of the norm of the wave vector 
\begin{equation}
  \label{eq:c0(k)}
  c_0 = c_0(k) \ ,
\end{equation}
which characterises the propagation of a water wave {\it without} accounting for air or wind. For example, $c_0 = \sqrt{\frac{g}{k} + \frac{\sigma \, k}{\rho\wat}}$ for gravito-capillary waves.
The wave growth coefficient $\gamma$ due to the effect of wind, according to Miles' approach,  can then be established from its equivalent in Miles' simplified reference conditions by applying a simple factor $\Xcoeff_0$:
\begin{equation}
  \label{eq:X0}
  \gamma = \Xcoeff_0 \times \gamma_{\sss M}
  \ .
\end{equation}
This coefficient $\Xcoeff_0$ captures all the hydrodynamics in the water domain, and can be obtained from the dispersion relation Eq.~(\ref{eq:c0(k)}) via straightforward algebra (see several examples in Table \ref{table:coefficients}).

It is important to underline that this correspondence is established without introducing further approximations.
Indeed, the procedure follows the same steps as those in previous publications \cite{Montalvo:2013a,Montalvo:2013b,Kharif:2020,Kern:2021,Chaubet:2024}, but lays them out in a generic way. 
The resulting method is rather general, and calculations are surprisingly easy, requiring no knowledge of hydrodynamics (several relevant examples are given in detail in \ref{app:framework}). Here we simply refer to Table \ref{table:coefficients} to highlight the multitude of phenomena which can be captured.

Using Table \ref{table:coefficients} and Eq.~(\ref{eq:beta:def}) we can deduce the amplitude growth coefficient (in a general hydrodynamic situation) from Miles' pressure coefficient $\beta$ (known from Eq.~(2.3) in Miles' seminal paper \cite{Miles:1957}) as
\begin{equation} 
  \frac{\gamma}{\omega} = \Xcoeff_0 \times \frac{s}{2} \, \frac{\beta}{\theta^2}
  \ . 
\end{equation}
Using expression (\ref{eq:fit:full}) for $\beta$  this can then be exploited readily.

\begin{table}
  \begin{tabular}{|c|c|c|c|c|} 
    \hline
    conditions & $c_0(k)$ & $\Hcoeff_0$ & $\Xcoeff_0$ & $\gamma$ \\ [0.5ex] 
    \hline
    Miles' ocean  & $\sqrt{\frac{g}{k}}$ & $1$ & $1$ & $\gamma_{\sss M}$  \\ 
    \hline
    water depth $h$ & $\sqrt{\frac{g}{k} \, {\mathbb T}}$ & $\sqrt{{\mathbb T}}$ & $\mathbb{T}$ & $\mathbb{T} \times\gamma_{\sss M}$ \\ 
    \hline
vorticity: shear rate $\Omega$ & $-\frac{\Omega}{2k} {\mathbb T} \pm \frac{1}{2}\sqrt{\frac{\Omega^2}{k^2}{\mathbb T}+\frac{4g}{k} {\mathbb T}}$
 & $\frac{1}{\sqrt{\frac{1}{\mathbb{T}} + \frac{\Omega}{k c_0}}}$ & $\frac{\mathbb{T}}{1 + \frac{1}{2} \, \frac{\Omega \, \mathbb{T}}{k c_0}}$     & $\frac{\mathbb{T}}{1 + \frac{1}{2} \, \frac{\Omega \, \mathbb{T}}{k c_0}} \times \gamma_{\sss M}$ \\ 
    \hline bulk viscosity (small) $\nu=\frac{\mu}{\rho\wat}$ & $\sqrt{\frac{g}{k}} - i \, 2 \nu k$ & 1 & 1 & $\gamma_{\sss M} - 2 \frac{k \nu}{\Re[c_0]}$ \\
    \hline
    surface tension $\sigma$ & $\sqrt{\frac{g}{k} + \frac{\sigma k}{\rho\wat}}$ & $\frac{1}{\sqrt{1-\frac{\sigma \, k}{\rho\wat \, c_0^2}}}$ & 1 & $\gamma_{\sss M}$ \\
    \hline
    general condition & $c_0(k)$ &$\frac{c_0(k)}{\sqrt{g/k}}$ & $\frac{\Hcoeff_0^2}{1-\frac{c_0}{\Hcoeff_0} \, \pder{\Hcoeff_0}{c_0}}$ & $\Xcoeff_0 \times \gamma_{\sss M}$ \\ [1ex] 
    \hline
  \end{tabular}
  %
  \caption{\label{table:coefficients}
    Key quantities ($\Hcoeff_0$, formula (\ref{eq:H0}) in \ref{app:framework},  and $\Xcoeff_0$, Eqs.~(\ref{eq:X0}) and (\ref{eq:chi0:f})) for applying the approach to various hydrodynamic conditions.
    For the sake of better readability we have abbreviated $\mathbb{T}=\tanh(kh)$ everywhere in this table.
    Details are provided in \ref{app:framework}. Note that $c_0(k)$ refers to the dispersion relation of the system where the air is replaced by a constant pressure field, which is therefore much simpler to establish than the full problem with air and wind.
  }
\end{table}

\section{Discussion and conclusions}

Miles' approach to wind-wave interaction is to this date the most convincing way to provide a prediction which entirely resides in first-principle physics-based mechanisms, rather than ad-hoc phenomenological assumptions.
However, the full complexity of combining hydrodynamics in wind and air, coupled by a free interface, makes this approach appear rather unwieldy.
Nevertheless, the main arguments and results, which we have summarised, are straightforward to understand in their spirit. Numerical work is only required once for a specific wind profile, and is no longer necessary at all for the commonly used logarithmic wind profile: a convenient fit provides an easy-to-use result for the growth coefficients, which fits Miles' result for all wind strengths for which the approach is expected to be valid.
Furthermore, although the result by Miles is often said to underestimate experimental results by a factor of the order of two, it yields in fact accurate results once the entire spectrum of waves is accounted for.
Finally, we have shown how additional features in the hydrodynamics may be incorporated in a very straightforward way, requiring very little mathematics and no hydrodynamics, based only on the dispersion relation of a wave propagating at the interface of the reference system without wind nor air.

To summarise, we have summarised the gist of Miles' approach and shown that it works. Furthermore, we have provided an easy-to-use formula rendering the main result, and vastly extended its domain of applicability by incorporating additional hydrodynamic parameters such as water depth, surface tension or water currents, requiring very little mathematics and no additional numerical effort.

We therefore think that the approach has reached a degree of maturity which removes all barriers for its use in practice, such as modelling or analysing experimental and observational data. Despite many remaining assumptions - and in particular a restriction to small wave amplitudes - the potential of the approach is now exposed, showing also the extent to which it can provide a backbone of understanding, onto which further extensions and corrections may then be added. We certainly hope that the  picture put forward by Miles may find a more prominent place in analysing wind-wave interactions than has been the case to date.

\section*{Disclosure Statement}
The authors are not aware of any affiliations, memberships, funding, or financial holdings that might be perceived as affecting the objectivity of this review. 

\section*{Acknowledgements}
This work was supported by the program HYDRAUMATH by the French government, managed by the National Research Agency (ANR) under the France 2030 initiative, grant number ANR-23-EXMA-0007. This work was also supported by the French State under the PPR FUTURISKS project 'Un océan de solutions', managed by the National Research Agency, under France 2030 (grant No. ANR-22-POCE-0002).

\bibliographystyle{elsarticle-num}

\bibliography{plea}

\clearpage
\appendix

\section{Miles beyond Miles: extending hydrodynamics is painless}
\label{app:framework}

\subsection{A new approach: an executive summary}

In a nutshell, the approach is as follows:
Miles has characterised the impact of wind on wave growth through a hydrodynamic analysis, which has been found to be well suited to the second stage of wave growth (once perturbations are initiated due to Phillips' mechanism). During this stage, perturbations are already present, and  grow exponentially. Miles has shown that the amplitude growth coefficient can be expressed as an integral which accounts for the energy transfer from the wind to the wave. In Miles' reference conditions (an infinitely deep ocean of incompressible, inviscid water, without any submarine currents) this leads to a single coefficient $\gamma$ for the (exponential) amplitude growth, which is a function of the wave age only. Recent work \cite{arxiv} has shown that this approach can be extended to more complex hydrodynamic conditions. Indeed, the coefficient $\beta$ characterising the impact of wind remains unchanged, but the energy transfer is modulated by another coefficient, the hydrodynamic factor $\Xcoeff_0$.

To understand why this works note that, whatever the hydrodynamics in the water domain, the perturbation in the air domain can depend only on the shape of the interface deformation (wavelength $k$ and amplitude $\eta$) and its phase velocity (the latter being linked to $k$   by the dispersion relation $c(k)$ appropriate to those conditions).
One can then show that, to the dominant order in $s=\rho\air/\rho\wat$, everything follows from the wave celerity $c_0(k)$ for a wave propagating in a simplified system where the air is absent, replaced by a constant pressure field \cite{arxiv}. This reflects the fact that the wave age $\theta_0$ remains the central parameter to the theory, and is therefore the key parameter to the amplitude growth coefficient $\gamma$. Assuming that the roughness length obeys the relation due to Charnock,
\begin{equation}
  \label{eq:z0:general}
    k \, \zo
    =  \frac{\Omega_{\sss CH}}{\theta_0^2}
  \ ,
\end{equation}
this generalises Eq.~(\ref{eq:charnock}) beyond Miles' reference conditions. One can then show \cite{arxiv} that the impact of modifying hydrodynamics can be accounted for through the coefficient $\Xcoeff_0$, which depends on the hydrodynamic conditions in the water as well as the wave age $\theta_0=c_0/{\mathcal U}$, but {\it not} on the wind as such. Furthermore, this coefficient can be calculated rather easily, directly from the dispersion relation of a wave propagating at the water surface {\it without air} (subject to a constant air pressure field), and thus {\it a fortiori} without wind. This is of course a significant simplification. Essentially, in order to account for the hydrodynamics, it then suffices to multiply  Miles' amplitude growth rate by the hydrodynamic factor $\Xcoeff_0$.

\subsection{The procedure, step by step}

Here we provide a pragmatic, operational summary of the steps to be carried out:
\begin{itemize}
\item
  Characterise the hydrodynamic conditions in the water domain (such as a finite water depth, a viscosity, etc) in terms of non-dimensional variables $\tilde{h}_1$, $\tilde{h}_2$.
  \\
  Note: these are proper hydrodynamic parameters, specific to the water domain. There is no question of wind yet, and therefore these dimensionless parameters should not use the wind strength or the air density. The wave sets a natural velocity scale as $c_0$ and a natural length scale as $1/k$, and thus a time scale $1/(k c_0)$.
\item
  Establish the dispersion relation $c_0=c_0(k)$ for a surface wave propagating on the simplified airless (and thus windless) system in which the air domain is replaced by a constant air pressure.
  \\
  Note: In many cases, this is already available in the literature.
\item
  Define the function
  \begin{equation}
    \label{eq:H0}
    \Hcoeff_0(c_0,...)
    := \frac{c_0}{\sqrt{g/k}}
    \ .
  \end{equation}
  Note: this coefficient is understood to be a function of all hydrodynamic parameters (water depth, shear rate, viscosity etc) comprised in the model, as well as (in some cases, see below) $c_0$ itself. It must {\it not}, however, depend on the reference velocity in Miles' system, $c_{0,M}$: if such a dependence were to occur, $\sqrt{g/k}$ would need to be expressed in terms of $c_0$ and the other hydrodynamic parameters. This will become clear in the example on the surface tension below\footnote{Essentially, this avoids the question of having to calculate a derivative $\partial (g/k)/\partial c_0$, which would be exceedingly awkward to evaluate.}.
\item  
  Calculate the conversion factor $\Xcoeff_0$, which captures the effect of hydrodynamics on the amplitude growth coefficient, as
  \begin{equation}
  \label{eq:chi0:f}
 \Xcoeff_0
 :=
 \frac{\Hcoeff_0^2}{1 - \frac{c_{0}}{\Hcoeff_0} \, \pder{\Hcoeff_0}{c_0}}
 \ ,
  \end{equation}
  as prescribed  by Eq.~(75) in \cite{arxiv}.
  \\
  Note : we assume for now that $c_0$ is real, the case of dissipation will be discussed below.
\item
  The amplitude growth coefficient $\gamma$ can now be determined simply by applying the hydrodynamic factor $\Xcoeff_0$ to Miles' result, {\it i.e.}
  \begin{equation}
    \gamma = \Xcoeff_0 \times \gamma_{\sss M}
    \ .
  \end{equation}
  Note: the pressure coefficients $\beta$, which is frequently used to quantify wave growth, remains the one determined by Miles {\it at the actual wave age}, accounting for the real hydrodynamics.
\end{itemize}

Alternatively, one may also proceed from the expression for the pressure perturbation accompanying the wave in the simplified airless (and thus windless) system: see Eqs.~(32) and (43) in \cite{arxiv}. However, since this expression is less commonly available in the literature, the approach based on the dispersion relation may prove simpler to exploit in practice.

\subsection{Examples} 

We first recapitulate examples also given in \cite{arxiv}, but here we address them directly based on the dispersion relation. In addition we then present an expression to account for surface tension.

In practice, one may wish to plot $\gamma$ as a function of some parameter, say $k$ (for a given $h$ and a given ${\mathcal U})$ or another,  say as a function of ${\mathcal U}$ (for a given $h$ and a given $k$) or, more appropriate to a laboratory setting, as a function of $h$ (for a given $k$ and a given ${\mathcal U}$). This can be achieved via
\begin{equation}
  \gamma(\theta) = \Xcoeff_0 (\theta,...) \times \gamma_{\sss M}(\theta) \ .
\end{equation}
Here $\Xcoeff_0(\theta,...)$ suggests a dependence on the wave age but also on other parameters, which determine the hydrodynamics, and where $\gamma_{\sss M}$ is related to $\beta$ via Eq.~(\ref{eq:beta:def}).
Ultimately all hydrodynamic scenarios can thus be derived directly from $\beta(\theta)$ where $\theta$ is evaluated as $\theta = c_0/{\mathcal U} = {\mathcal H}_0 \, \sqrt{g/k} /{\mathcal U}$~: this is the pivotal point of the approach.

\subsubsection*{Example: finite depth}

The only non-dimensional hydrodynamic parameter is $\tilde{h}=kh$, and the dispersion relation with a constant air pressure field is known to be $c_0 = \sqrt{g/k} \, \sqrt{\tanh(\tilde h)}$.
We have
\begin{equation}
  \Hcoeff_0 = \sqrt{\tanh(kh)}
  \ .
\end{equation}
This does not depend on $c_0$, and therefore the derivative in the denominator of Eq.~(\ref{eq:chi0:f}) vanishes. Therefore we have simply
\begin{equation}
  \Xcoeff_0 = \Hcoeff_0^2 = \tanh(\tilde{h})
  \ ,
\end{equation}
and thus $\gamma = \tanh(kh) \times \gamma_{\sss M}$, as found in \cite{Montalvo:2013a,Montalvo:2013b,Montalvo:thesis:2014}.

\subsubsection*{Example: vorticity} 

The non-dimensional parameter characterising the shear flow is
$\tilde{\Omega}$, which is a non-dimensionalised vorticity.
Note that it depends on $c_0$ (since $\tilde{\Omega} = \frac{\Omega}{k \, c_0}$).
Using the dispersion relation, which is available in the literature \cite{Hur:2017}, yields
\begin{equation}
  \Hcoeff_0 = \Hcoeff_0(\tilde{T},\tilde{\Omega}) = \left(\frac{1}{\tilde{T}} + \tilde{\Omega}\right)^{-1/2}
  \ .
\end{equation}
The derivative 
\begin{equation}
  \pder{\Hcoeff_0(\tilde{\Omega})}{c_0}
  = \pder{\Hcoeff_0}{\tilde\Omega} \, \pder{\tilde\Omega}{c_0}
  = \frac{\tilde\Omega}{2 \,c_0} \, \Hcoeff_0^3
  \ ,
\end{equation}
leads to the hydrodynamic factor $\Xcoeff_0$ as
\begin{equation}
  \Xcoeff_0
  = \frac{1}{\frac{1}{\Hcoeff_0^2} - \frac{\tilde{\Omega}}{2}}
  = \frac{\tilde{T}}{1+\tilde\Omega \tilde{T}/2}
  \ ,
\end{equation}
where the last correspondence is obtained upon substituting $\Hcoeff_0^2$.
This reproduces the known results \cite{Chaubet:2024} (although some algebra is required to make contact with these results, see the Appendix of \cite{arxiv} for details).

\subsubsection*{Example: bulk viscosity}

When there is dissipation in the bulk, this already entails an imaginary contribution to the celerity $c_0$ which characterises the wave propagation, even in the absence of wind and air.
 Here it is sufficient to limit the discussion to the 'weakly viscous' case, characterised by $\nu k/c_0 \ll 1$: our approach relies on this condition (see the discussion in \cite{arxiv}), but this is also the relevant case for ocean waves.
In this case, there is a (negative) growth coefficient due to viscosity, which is to be added to the growth coefficient due to wind:
\begin{equation}
  \gamma_{tot} = \gamma_{visc} + \gamma_{wind}
\end{equation}
where $\gamma_{visc}=-2\nu k^2$ is a well-known expression. The contribution $\gamma_{wind}$ due to the wind can now be calculated as before, from Eqs.~(\ref{eq:H0}) and (\ref{eq:chi0:f}), but retaining only the real part to the celerity, $\Re[c_0]$.

The full dispersion relation for surface waves on a body of viscous water is reviewed in \cite{Dias:2008}, but in the condition of 'weak' viscosity it can be simplified (see \eg \cite{Dias:2008,Lamb:book}) to
\begin{equation}
  c_0\approx \sqrt{g/k} - 2i\nu k \,
\end{equation}
which therefore implies a (small) imaginary part.

Since the real part of $c_0$ is unchanged with respect to the expression given by Miles, this also holds for the real part of $\Hcoeff_0$, and therefore ultimately $\gamma_{wind} = \gamma_{wind,M}$, \ie the actual growth coefficient due to wind is unchanged (which is coherent, since viscosity has already been accounted for by adding in the (negative) contribution $\gamma_{visc}$):
\begin{equation}
\gamma_{tot} = \gamma_{\sss M}  - 2 \nu k^2 
  \ .
\end{equation}

\subsubsection*{Example: capillarity}

When a surface tension $\sigma$ is present, the dispersion relation for gravito-capillary waves propagating at the surface is known to be
\begin{equation}
  c_0 =  \sqrt{\frac{g}{k} + \frac{\sigma \, k}{\rho\wat}}
  \ .
\end{equation}
This implies that
\begin{equation}
  \frac{g}{k} = c_0^2 - \frac{\sigma k}{\rho\wat}
  \ ,
\end{equation}
which we can then use to express the coefficient $\Hcoeff_0$ as a function of $c_0$ as
\begin{equation}
  \Hcoeff_0
  = \frac{c_0}{\sqrt{g/k}}
  = \frac{1}{\sqrt{1 - \frac{\sigma \, k}{\rho\wat \, c_0^2}}}
  \ .
\end{equation}
Thus, from straightforward algebra,
\begin{equation}
  \frac{c_0}{\Hcoeff_0} \, \pder{\Hcoeff_0}{c_0}
  =
  - \frac{\sigma \, k}{\rho\wat \, c_0^2} \, \Hcoeff_0^2
  \ ,
\end{equation}
from which it follows directly, via Eq.~(\ref{eq:chi0:f}), that
\begin{equation}
\Xcoeff_0
=
\frac{\Hcoeff_0^2}{1 - \left(- \frac{\sigma \, k}{\rho\wat \, c_0^2} \, \Hcoeff_0^2\right)}
= \frac{1}{\frac{1}{\Hcoeff_0^2} +  \frac{\sigma \, k}{\rho\wat \, c_0^2} }
= 1
\ .
\end{equation}

The statement here is that, although the presence of the surface tension does affect the wave celerity, it does {\it not} alter the energy transfer from wind to wave if the comparison is made at the same wave age. Recall nevertheless that this demonstration excludes viscosity, currents, wind-induced surface flow, compressibility and many other potential complications. 

Although this result may appear surprising, it is in agreement with a much more general result established for the impact of surface tension, and deserving a more detailed discussion \cite{us:wip}.

\section{Growth coefficient averaged over a spectrum}
\label{app:averaged}

Here we argue that the correction due to the presence of a complete wave spectrum, rather than just an individual frequency, makes the predictions by Miles coincide with observational data. To do so, we focus on the linear regime in the energy growth coefficient
\begin{equation}
  \Gamma(\theta)/\omega = 2 \gamma(\theta)/\omega
  \ ,
\end{equation}
see Fig.~\ref{fig:realdata}, and show that the spectrally averaged growth coefficient differs by a simple numerical factor, which we determine explicitly for the case of the simplified hydrodynamic conditions.
This linear range, say $10^{-1} < 1/\theta^* < 1$, essentially corresponds to the plateau region. We therefore have $\beta(\theta)\approx const$ in Fig.~\ref{fig:fit:full}. Alternatively, via Eq.~(\ref{eq:beta:def}), this means that $\Gamma/\omega \sim\theta^{-2}$. 

We now need to evaluate the integral Eq.~(\ref{eq:spectrum:average})
using the spectrum by Pierson-Moskovitz from Eq.~(\ref{eq:spectrum:PM}):
 \begin{equation}
  \label{eq:gamma:moyen:def}
  \avg{\frac{\Gamma}{\omega}}
  = \frac{\int_{0}^{\infty} \frac{\Gamma(\omega)}{\omega} \, S(\omega) \, d\omega}{\int_{0}^{\infty} S(\omega)  \, d\omega}
  \ .
\end{equation}
To this end we link the inverse frequency to the wave age through
\begin{equation}
  \theta= \frac{c}{U}=\sqrt{\frac{g}{k} \frac{1}{U}}=\frac{g}{U}\frac{1}{\sqrt{gk}}=\frac{g}{U}\frac{1}{\omega}
  \ ,
\end{equation}
which implies a direct proportionality, $\Gamma/\omega =  A \,\omega^2$, with some constant $A$. Thus
\begin{equation}
  \avg{\frac{\Gamma}{\omega}}
  =
  \frac{
    \int_0^\infty A \, e^{-\frac{5}{4} \, \left(\frac{\omega_{p}}{\omega}\right)^4}\omega^{-3} \, d\omega}{%
    \int_0^\infty  e^{-\frac{5}{4} \, \left(\frac{\omega_{p}}{\omega}\right)^4} \, \omega^{-5} \, d\omega
  }
  =
  A \, \omega_p^2  \, \sqrt{\frac{5}{4}} \,
  \frac{%
    \int_0^{\infty}  e^{-\xi^2} \,  d\xi
  }{%
    \int_0^{\infty} \xi \, e^{-\xi^2} \,  d\xi
  }
  =
  A \ \sqrt{\frac{5}{4}} \, \omega_p^2  \
  \frac{%
    \sqrt{\pi}/2
  }{%
    1/2
  }
  = 
  \sqrt{\frac{5 \pi}{4}} \, A \, \omega_{p}^2    
  \ ,
\end{equation}
where the substitution $\xi= \sqrt{5/4} \, (\omega_{p}/\omega)^2$ has been made. The result in Eq.~(\ref{eq:gamma:moyen:def}) has to be compared to the expression $\Gamma_p/\omega_p := A \, \omega_{p}^2$ for a monochromatic wave (the peak of which thus corresponds, trivially, to the only frequency present).
Therefore, for the energy growth coefficient
\begin{equation}
  \label{eq:Gamma}
  \avg{\frac{\Gamma}{\omega}} = \sqrt{\frac{5 \pi}{4}} \  \frac{\Gamma_p}{\omega_p} 
  \, .
\end{equation}
This means that,  when using only the peak frequency, rather than considering the entire spectrum, the observed growth coefficients will be underestimated by a factor of  $\sqrt{5\pi/4} \approx 1.98$, which results in the corresponding shift on the logarithmic scale in Fig.~\ref{fig:realdata}.

\section{Deconstructing Figure~\ref{fig:realdata}: data and fits}
\label{app:log4}

Some of the elements appearing in Fig.~\ref{fig:realdata}, although in principle straightforward, may be difficult to determine explicitly, due to competing definitions by various authors or in different fields, as well as what appears to be typographical errors in some published formulas and other mistakes. Here we report a complete explanation on how we have constructed the plotted quantities,  indicating specifically the possible pitfalls when attempting to do so. Hopefully this might be useful to the community, and avoid spending time on trial and error.

The data points represented in Fig.~\ref{fig:realdata} have been taken from Plant's  \cite{Plant:1982} compilation of observational as well as laboratory data produced by various authors. Symbols have been preserved in order to link back to the original data of each source in this compilation. Similar (but not strictly identical) data sets have been plotted in other publications (Fig.~1 in \cite{Janssen:1991}, Fig.~1 in \cite{VanDuin:1992},  Fig.~3.3 in \cite{Janssen:book}), all of which refer back to \cite{Plant:1982}.

The fine gray line labelled {\it Miles(monochromatic)}  refers to the energy growth coefficient $\Gamma = 2 \times \gamma$, introduced by Miles \cite{Miles:1957}, where the amplitude growth coefficient $\gamma$ has been constructed from Eq.~(\ref{eq:beta:def}) and the fit function established in Eq.~(\ref{eq:fit:full}). This directly leads to
\begin{equation}
  \frac{\Gamma_{\sss M}}{2 \pi  f} = 2 \times \frac{\gamma_{\sss M}}{\omega}
  = 2 \times \frac{s}{2} \, \frac{\beta(\theta)}{\theta^2}
  \ ,
\end{equation}
where $\omega = 2 \pi \, f$ and the  {\it wave age} $\theta$ is understood to be
\begin{equation}
  \theta = \frac{c}{{\mathcal U}}
  \ ,
\end{equation}
in keeping with the notation used by Miles.

In other sources, and also in Plant \cite{Plant:1982}, the ratio $c/u^*$ is used instead to define the wave age, and we distinguish this quantity by adding a star. Since the identification $u^* = \kappa \, {\mathcal U}$ can be made via Eq.~(\ref{eq:windprofile:log}), we thus have
\begin{equation}
  \frac{1}{\theta} = \frac{1}{\kappa} \, \frac{1}{\theta^*}
  \ .
\end{equation}
Putting it all together as a function of $\frac{1}{\theta^*}$ we thus have
\begin{equation}
  \frac{\Gamma_{\sss M}}{f} =  2 \pi \, s \,   \frac{\beta(\frac{\kappa}{1/ \theta^*})}{  \left( \frac{\kappa}{1/\theta^*}\right)^2}
  \ ,
\end{equation}
which is what we have plotted in Fig.~\ref{fig:realdata}, without any ambiguity.

The last graph, the fine red line, is that of a previously proposed ad-hoc analytical function, apparently constructed in \cite{Janssen:1991} to describe the data in Fig.~\ref{fig:realdata}. For this, more care is required.
Note first that Eq.~(19) in \cite{Janssen:1991} states that
\begin{equation}
  \frac{\gamma_{\sss M}}{\omega} = s \, \beta \, \left(\frac{u^*}{c}\right)^2
  \ ,  
\end{equation}
which deviates from the definition used above (and dating back to Miles). In order to make contact with our definitions several points have to be remembered: (i) this definition is based on the alternative definition of the wave age, $\theta^* = \theta/\kappa $, already mentioned above, and this implies that $\beta$ in \cite{Janssen:1991} corresponds to our $\beta / \kappa^2$;  (ii) since the prefactor is $s$, rather than $s/2$, this implies that this coefficient is in fact {\it not} the amplitude growth coefficient but, rather, the energy growth coefficient $\Gamma$; (iii) the expression used for $\beta$ is {\it not} the one we have derived above, but an ad-hoc fit stated just below (Eq.~(\ref{eq:beta:janssen})), which we now construct in detail.

Using the functions and coefficients introduced in \cite{Janssen:1991}, and also reproduced in \cite{Janssen:book}, we have
\begin{eqnarray}
  \mu = \left(\frac{u^*}{\kappa \, c}\right)^2 \, \Omega \, e^{\kappa \, c/{\mathcal U}^*} &&\qquad(\mbox{Eq. (16) in \cite{Janssen:1991})}
  \\
  \Omega = g \, \zo \, \left(\frac{\kappa}{u^*}\right)^2
  &&\qquad(\mbox{Eq. (17) in \cite{Janssen:1991}, with $z_1=0$)}
  \\
  \beta = \frac{1.2}{\kappa^2} \, \mu \, \ln^4 \mu
  &&\qquad(\mbox{Eq. (18) in \cite{Janssen:1991})}                                                               \label{eq:beta:janssen}
  \\
  \zo = \alpha \, \frac{\left(u^*\right)^2}{g}
  &&\qquad(\mbox{Eq. (13) in \cite{Janssen:1991})}
  \ .
\end{eqnarray}

From these expressions it is straightforward to deduce that
\begin{equation}
  \frac{\Omega}{\kappa^2} = \alpha_{\sss CH}
  \ ,
\end{equation}
which then leads to an expression for $\mu$ in terms of $\theta^*$:
\begin{equation}
  \mu = \frac{\alpha_{\sss CH} \, e^{\kappa \, \theta^*}}{\left(\theta^*\right)^2}
          \ .
\end{equation}
For any given wave age $\theta$ (or, here, $\theta^*$) this can be used to calculate the value of $\mu$, from which the value of $\beta$ follows directly.
Taking into account remark (ii) made above, this ultimately yields
\begin{equation}
  \label{eq:janssen:fit}
  \frac{\Gamma_{\sss M}}{f}
  =
  2 \pi \, s \, \frac{1.2}{\kappa^2} \, \mu \,  \frac{\ln^4\mu}{\left(\theta^*\right)^2}
  \ .
\end{equation}
This is the expression we have used to plot Janssen's ad-hoc function in Fig.~\ref{fig:realdata}.
\\

Finally, it appears useful to further point out two additional observations which might reveal confusions or inconsistencies.
First,  in Fig.~2 in \cite{Janssen:1991} it is indicated that this is the quantity which is plotted, given by Eq.~(\ref{eq:janssen:fit}) (although there is some ambiguity as to what the 'straight line' refers to, since there are several straight lines). However, the graph which would correspond to Eq.~(\ref{eq:janssen:fit}) appears to be straight in the central region of $1/\theta^*$, whereas our plot, of this same function, produces a slight dip in this region.
Second, a very similar graph is produced in Fig.~3.3 in \cite{Janssen:book}, but is referred to as {\it Miles' theory} (although Miles' predictions are generally stated to fall below the data).

\end{document}